\documentclass[aps,preprint,floatfix,amssymb,amsmath,prb,endfloats]{revtex4}
\usepackage{graphicx}
\begin{document}
\title{
Inherent Flexibility and Protein Function: The Open/Closed Conformational Transition 
in the N-Terminal Domain of Calmodulin 
}

\author{Swarnendu Tripathi and John J. Portman}

\affiliation{
Department of Physics, Kent State University, Kent, OH 44240
}

\begin{widetext}
\begin{abstract} 
The key to understanding a protein's function often lies in its
conformational dynamics. We develop a coarse-grained variational model
to investigate the interplay between structural transitions,
conformational flexibility and function of N-terminal calmodulin
(nCaM) domain. In this model, two energy basins corresponding to the
``closed'' apo conformation and ``open'' holo conformation of nCaM
domain are connected by a uniform interpolation parameter. The
resulting detailed transition route from our model is largely
consistent with the recently proposed EF$\beta$-scaffold mechanism in
EF-hand family proteins. We find that the N-terminal part in calcium
binding loops I and II shows higher flexibility than the C-terminal
part which form this EF$\beta$-scaffold structure. The structural
transition of binding loops I and II are compared in detail. Our model
predicts that binding loop II, with higher flexibility and early
structural change than binding loop I, dominates the conformational
transition in nCaM domain.
\end{abstract}  
\end{widetext}                                                               

\date{November 4, 2007} 

\maketitle


\clearpage

\section*{INTRODUCTION}

Many protein functions fundamentally depend on structural flexibility. Complex
conformational transitions, induced by ligand binding for example, are
often essential to proteins participating in regulatory networks or
enzyme catalysis. More generally, a protein's ability to sample a
variety of conformational sub-states implies that proteins have an
intrinsic flexibility and mobility that influences their
function.\cite{frauenfelder:wolynes:91,gerstein:chothia:94} While
experimental measurement can offer direct dynamical information about
specific residues, uncovering the detailed mechanisms controlling
conformational transitions between two meta-stable states is often
elusive.  In this paper we present an analytic model that aims to
clarify the relationship between main-chain dynamics and the
mechanisms controlling conformational transitions of flexible
proteins. In particular, we examine the mechanism for the open/closed
transition of the N-terminal domain of Calmodulin (nCaM) to explore
how calcium binding and target recognition can be understood by
changes in the mobility and the degree of partial order of the protein
backbone.

\begin{figure}
   \begin{center}
      \includegraphics[width=4.5in]{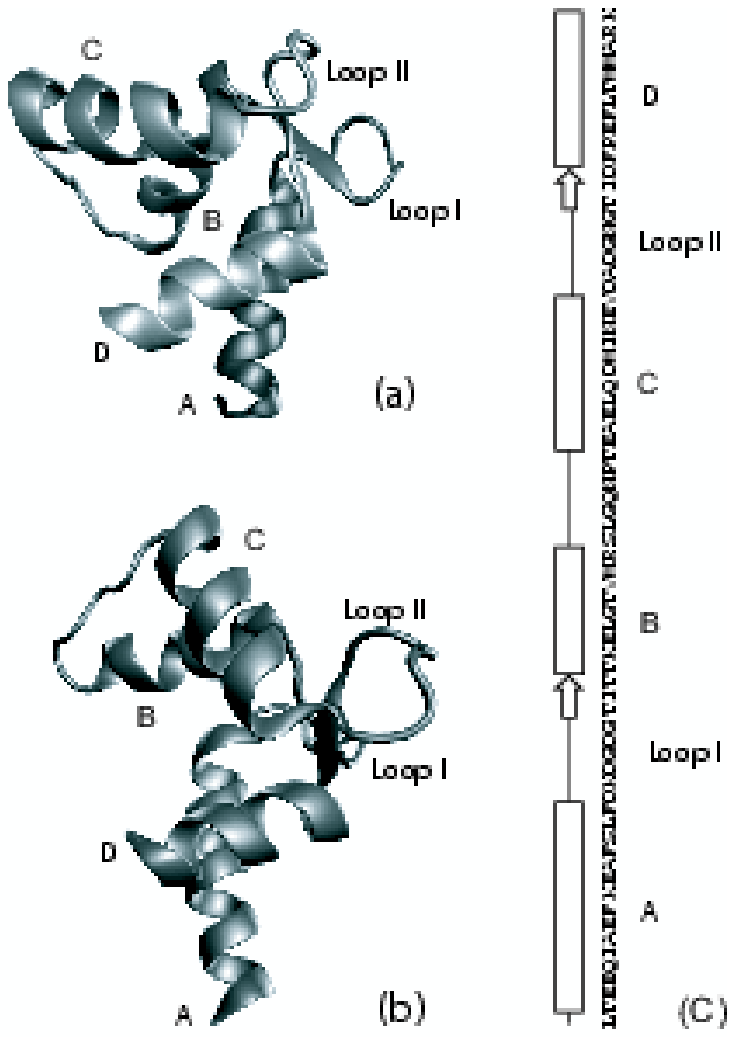}
      \caption{}
      \label{fig:2nd_3D}
   \end{center}
\end{figure}

Calmodulin (CaM) may be an ideal model system to illustrate how
conformational flexibility is a major determinant of biological
function. CaM is found in all eucaryotic cells and functions as a
multipurpose intracellular Ca$^{2+}$ receptor, mediating many
Ca$^{2+}$-regulated processes. CaM is a small (148 amino acid)
dumbbell shaped protein with two domains connected by a flexible
linker. Each domain
of CaM contains a pair of helix-loop-helix Ca$^{2+}$ -binding motifs
called EF-hands (helices A/B and C/D in the N-terminal domain). These
two EF-hands are connected by a flexible B/C helix-linker (see
Fig.~\ref{fig:2nd_3D}). In each domain the four helices of apo-CaM are
directed in a somewhat antiparallel fashion giving the domains a
relatively compact structure while leaving the Ca$^{2+}$-binding loops
exposed. The conformational change induced by binding Ca$^{2+}$ can be
described as a change in EF-hand interhelical angle (between helices
A/B and C/D) from nearly antiparallel (apo, closed conformation) to
nearly perpendicular (holo, open conformation) orientation.
Further this domain opening mechanism in nCaM
indicates that binding of Ca$^{2+}$ occurs almost exclusively within
EF-hands, not between them.\cite{nelson:chazin:98} The structural
rearrangement from closed to open exposes a large hydrophobic surface
rich in Methionine residues responisble for molecular recognition of 
various cellular targets such as myosin light chain kinase.

The high flexibility of CaM is essential to its function.
The flexibility of the central helix linking the two
domains allows the activated domains to simultaneously interact with target peptides.
The conformational flexibility of the domains themselves allow for
considerable binding promiscuity of target peptides, a property
essential to its function as a primary messenger in Ca$^{2+}$ signal
transduction.\cite{brokx:makhatadze:01,chou:bax:01} While similar in
structure and fold, the two domains of CaM are quite different in
terms of their flexibility, melting temperatures, and Ca$^{2+}$-binding
affinities.\cite{tsalkova:privalov:85,linse:forsen:91}

The conformational dynamics of Ca$^{2+}$-loaded and Ca$^{2+}$-free
CaM are well characterized by solution 
NMR.\cite{chou:bax:01,baber:tjandra:01}
Site specific internal dynamics monitored by model
free order parameters $S^2$, indicate that the
helices of the apo-CaM domains are well-folded on the picosecond to
nanosecond timescale, while the Ca$^{2+}$-binding loops, helix-linker
and termini are more flexible.\cite{malmendal:akke:99}
On the other hand, spin-spin relaxation (or transverse
auto-relaxation) rates, $R_2$, 
indicate that the free and bound forms of the regulatory protein
exchange on the millisecond timescale.\cite{ishima:torchia:00}
Akke and coworkers have investigated the rate of conformational exchange
between the open and closed conformational substrates of C-terminal CaM
(cCaM) domain by NMR $^{15}$N spin relaxation
experiments.\cite{evenas:akke:99}
Comparison of exchange rates as a function of Ca$^{2+}$ concentration
have established that the conformational exchange in apo-cCaM
involves an equilibrium switching between the closed and open states that
is independent of Ca$^{2+}$ concentration.\cite{malmendal:akke:99}

X-ray crystallography temperature factors give additional insight into
the conformational freedom and internal flexibility of CaM in the open
and closed state. Recently, Grabarek proposed a detailed mechanism of
Ca$^{2+}$ driven conformational change in EF-hand proteins based on
the analysis of a trapped intermediate X-ray structure of
Ca$^{2+}$-bound CaM mutant.\cite{grabarek:05} This two-step Ca$^{2+}$-binding
mechanism is based on the hypothesis that Ca$^{2+}$-binding and the
resultant conformational change in all two EF-hand domains is
determined by a segment of the structure that remains fixed as the
domain opens. This segment, called the EF-hand-$\beta$-scaffold, 
refers to the bond network that connects the two Ca$^{2+}$ ions. It includes 
the backbone and the two hydrogen bonds formed by the residues in the 8$^{th}$ 
position of binding loops (Ile27 and Ile63) and the C=O groups of the residues 
in the 7$^{th}$ position of the binding loops (Thr26 and Thr62).\cite{grabarek:06} 
Indeed, in the absence of Ca$^{2+}$, the
N-terminal end of the binding loop is found to be poorly structured
and very dynamic from NMR 
structures~\cite{evenas:akke:99,kuboniwa:bax:95,zhang:ikura:95} and 
X-ray temperature factors.\cite{grabarek:05} Functional distinction between the
two ends of the binding loops in the domain opening mechanism is
buttressed by the great variability of the amino acid sequences of the
N-terminal ends of the Ca$^{2+}$-binding loops compared with the
more conserved C-terminal ends across a variety of different EF-hand
Ca$^{2+}$-binding proteins.\cite{grabarek:06}

In this paper, we study the role of flexibility in the conformational
transition of CaM through an extension of a coarse-grained variational model
developed to characterize protein
folding.\cite{portman:wolynes:98,portman:wolynes:01a,portman:wolynes:01b}
This model accommodates two meta-stable folded conformations as minima of the
calculated free energy surface. The natural order parameters of this model, discussed
in detail in the methods section, is well suited to describe partially
ordered ensembles essential to the conformational dynamics of flexible
proteins. Transition routes and conformational changes of the protein
are determined by constrained minimization of a variational free energy
surface parameterized by the degree of localization of each residue
about its mean position. The computational time to calculate the transition
route for nCaM is on the order of several minutes on a typical single-processor
PC.

In addition to extensive experimental work characterizing the inherent
flexibility of CaM, our results also benefit from all atom molecular
dynamics simulations~\cite{wriggers:schulten:98,vigil:garcia:01} as
well as recent coarse-grained simulations inspired by models developed
to characterize protein folding.\cite{zuckerman:04,gwei:hummer:07}
Although subject to systematic errors due to approximations, analytic
models have the important advantage that the results are free of
statistical noise that can obscure simulation results (particularly
troublesome when characterizing low probability states).

\section*{MODEL AND METHODS}
 
A configuration of a protein is
expressed by the $N$ position vectors of the $\alpha$-carbons of the
polypepetide backbone. We are interested in describing transitions between two
known structures denoted by $\{\mathbf{r}_i^{N_1}\}$ and
$\{\mathbf{r}_i^{N_2}\}$. Partially ordered
ensembles of polymer configurations are described by a reference
Hamiltonian
\begin{equation} \label{eq:model}
  \mathcal{H}_0/k_{\mathrm{B}}T = \frac{3}{2a^2} \sum_{ij}\mathbf{r}_i\Gamma_{ij}\mathbf{r}_j 
  + \frac{3}{2a^2} \sum_{i}C_i[\mathbf{r}_i - \mathbf{r}_i^{N}(\alpha_i)]^2.        
\end{equation}
where $T$ is the temperature and $k_{\mathrm{B}}$ is Boltzmann's constant.
Here, the first term enforces chain connectivity, in which
the connectivity matrix, $\Gamma_{ij}$, corresponds to 
a freely rotating chain with mean bond length $a = 3.8$\AA and 
valance angle between successive bond vectors set to
by $\cos \theta = 0.8$.\cite{bixon:zwanzig:78} The $N$ variational
parameters, $\{C\}$, control the magnitude of the fluctuations about
$\alpha$-carbon position vectors $\mathbf{r}_i^{N}(\alpha_i) =
\alpha_i\mathbf{r}_i^{N_1} + (1 - \alpha_i)\mathbf{r}_i^{N_2}$. The
$N$ variational parameters, $\{\alpha\}$ ($0 \le \alpha_i \le 1$),
specify residue positions as an interpolation
between 
$\{\mathbf{r}_i^{N_1}\}$ to
$\{\mathbf{r}_i^{N_2}\}$. 

The Boltzmann weight for a constrained chain described by $\mathcal{H}_0$ is proportional to
\begin{equation}
\omega(\{C\},\{\alpha\}) \propto \exp \left[ -\frac{3}{2a^2}
\sum_{ij} (\mathbf{r}_i - \mathbf{s}_i)  G_{ij}^{-1} (\mathbf{r}_j - \mathbf{s}_j) \right]
\end{equation}
where $G_{ij}$ denotes the correlations of monomers $i$ and $j$
relative to the mean locations,
$G_{ij} = \langle \delta \mathbf{r}_i \cdot \delta \mathbf{r}_j
\rangle_0/a^2$ with $\delta \mathbf{r}_i = \mathbf{r}_i -
\mathbf{s}_i$. Here, the correlations $G_{ij}$ are given by the matrix
inverse
$G_{ij}^{-1} = \Gamma_{ij} +
C_i\delta_{ij}$, and the mean positions of each monomer 
$\mathbf{s}_i = \sum_{j}G_{ij}C_j \mathbf{r}_j^{N}(\alpha_j)$ interpolate between
the coordinates in each native structure,
\begin{equation} \label{eq:mean position}
  \mathbf{s}_i = \sum_{j}G_{ij}C_j[\alpha_j\mathbf{r}_j^{N_1} + (1 -
\alpha_j)\mathbf{r}_j^{N_2}].
\end{equation}
The statistical properties of a structural ensemble
can be described in terms of the first two moments $\mathbf{s}_i$ and
$G_{i,j}$ since $\mathcal{H}_0$ is harmonic. 

In this model, the probability for a particular configurational ensemble at temperature
$T$ is given by
the variational free energy $F(\{C\},\{\alpha\}) = E(\{C\},\{\alpha\}) -
TS(\{C\},\{\alpha\})$. Here,  $S(\{C\},\{\alpha\})$ is
the entropy loss due to localizing the residues around the mean postions
\begin{equation}
S(\{C\},\{\alpha\})/k_{\mathrm{B}} = \frac{3}{2}\log \det G -\frac{3}{2a^2} \sum \mathbf{s}_i \Gamma_{ij} \mathbf{s}_j 
+ \frac{3}{2}\sum C_i G_{ii} .
\end{equation}
The energy is derived from two-body interactions between native
contacts, $E(\{C\},\{\alpha\}) = \sum_{[i,j]} \epsilon_{ij} u_{ij}$,
where $u_{ij}$ 
is the average of the pair potential $u(r_{ij})$ over $\mathcal{H}_0$, and 
$\epsilon_{ij}$ is the strength of a fully formed contact between residues $i$ and $j$ given
by Miyazawa-Jernigan interaction parameters.\cite{miyazawa:jernigan:96}
The sum is restricted to a set of contacts determined by pairs of
residues in the proximity in each of the meta-stable
conformations. The pair potential between two monomers is developed by 
a sum of three Gaussians
$u(r) =  \gamma_{\mathrm{s}}
e^{-3\beta_\mathrm{s} r^2 /2a^2 } + \gamma_{\mathrm{i}}
e^{-3\beta_\mathrm{i} r^2 /2a^2 } -\gamma_{\mathrm{l}}
e^{-3\beta_\mathrm{l} r^2 /2a^2 }$.
The parameters are chosen so that $u(r)$ has a minimum 
at $r^* = 1.6 a$ with value $u_{ij}(r^*) = -1$ formed
by the long-range attractive interactions
$(\gamma_{\mathrm{l}} = 6.0, \beta_{\mathrm{l}} = 0.27)$ and
intermediate-range repulsive interaction $(\gamma_{\mathrm{i}} = 9.0,
\beta_{\mathrm{l}} = 0.54)$ as in Ref.~\onlinecite{portman:wolynes:01a}. Excluded
volume interactions are represented by a short-range repulsive potential
with $\beta_{\mathrm{s}} = 3.0$ and $\gamma_{\mathrm{s}}$ is chosen so that each
contact has $u_{ij}(0)/\epsilon_0 = 100$, where $\epsilon_0$ is the basic energy
unit of the Miyazawa-Jernigan scaled contacts.\cite{miyazawa:jernigan:96}
The energy of
a contact between residues $i$ and $j$ in a partially ordered chain is given by
\begin{eqnarray}\label{uij}
\epsilon_{ij} u_{ij} &=& \epsilon_{ij}\langle u(r_{ij}) \rangle_0 \\ \nonumber
&=& \epsilon_{ij}  \sum_{\mathrm{k} =
(\mathrm{s,i,l})}\frac{\gamma_\mathrm{k}}{(1 +
\beta_{\mathrm{k}}\delta G_{ij})^{3/2}} \exp \left[ -\frac{3}{2a^2}
\frac{(\mathbf{s}_{i}-\mathbf{s}_j)^2}{1 + \beta_k \delta G_{ij}} \right].
\end{eqnarray}

In this work, we consider a two-state model in which
the contacts are separated into three sets:
$(i)$ contacts that occur in reference structure (1) only, $(ii)$
contacts that occur in reference structure (2) only, and $(iii)$
contacts in common from both reference structures. Then, we consider
that each contact involved exclusively with only one structure is in
equilibrium with energy from the other state (which is zero). That is,
we replace the pair energy for contacts in sets $(i)$ and $(ii)$ according to
\begin{equation}\label{eq:contact}
\epsilon_{ij} u_{ij} = 
-k_B T \log \left[ 1 + \exp(-\epsilon_{ij}\langle u(r_{ij})\rangle_0/k_BT) \right]. 
\end{equation}
This form is analogous to coupling between conformational basins in
folding-inspired molecular dynamics
simulation.\cite{maragakis:karplus:05,best:hummer:05,okazaki:wolynes:06} 
Contacts described by
Eq.~\ref{eq:contact} independently switch on or off depending on the
conformational density characterized by a set of constraints $\{C,\alpha\}$.

Analysis of the free energy surface parameterized by $\{C,\alpha\}$
follows the program developed to describe
folding:\cite{portman:wolynes:01a} the ensemble of structures
controlling the transition is characterized by the monomer density at
the saddlepoints of the free energy. At this point, we simplify our
model and restrict the interpolation parameter $\alpha_i$ to be the
same for all residues, $\alpha_i = \alpha_0$ following Kim et
al..\cite{kim:chirikjian:02} Then, the numerical problem simplifies to
minimizing the free energy with respect to $\{C\}$ rather than finding
saddlepoints in $\{C,\alpha\}$.

To explore the nature of conformational dynamics in detail, we apply
this model to the N-terminal domain of CaM (nCaM). In particular, we
use residues numbered 4-75 of unbound nCaM (apo, 1cfd) and bound nCaM
(holo, 1cll) (see Fig.~\ref{fig:2nd_3D}). In our model, we have
defined closed nCaM (1cfd) as structure (1) and open nCaM (1cll) as
structure (2).  Thus, the interpolation parameter $\alpha_0 = 1$
corresponds to the closed state, and $\alpha_0 = 0$ corresponds to the
open state.  The coordinates of the open/closed structure was rotated
to minimize the rmsd of $\alpha$-carbons between the two
structures.\cite{russell:barton:92} We note 
global alignment has the risk of possibly obscuring or averaging out some 
local structural differences. The temperature $T$ for the 
open/closed transition is taken to be the folding temperature
($T_\mathrm{f}$) of the open (holo, 1cll) structure with
$k_\mathrm{B}T_\mathrm{f}=2.0$. For comparison, the folding
temperature for closed (apo, 1cfd) structure is
$k_\mathrm{B}T_\mathrm{f}=1.9$.

For a given set of constraints, $\{C,\alpha\}$, the
monomer density of a partially ordered ensemble can be characterized
by the Gaussian measure of similarity to conformation described by
$\{\mathbf{r}_i^{N_1}\}$
\begin{eqnarray}
  \rho_i^{(1)}[\{C,\alpha\}] &=& \left\langle 
  \exp\left[ -\frac{3\alpha^N}{2a^2}(\mathbf{r}_i - \mathbf{r}_i^{N_1})^2\right]
  \right\rangle_0 \nonumber\\
  \label{eq:density 1}
  &=& (1 + \alpha^NG_{ii})^{-3/2}\exp\left[ -\frac{3\alpha^N}{2a^2}
    \frac{(\mathbf{s}_i - \mathbf{r}_i^{N_1})^2}{1 + \alpha^NG_{ii}}\right].
\end{eqnarray} 
Similarly, the structural similarity to the conformation described
by $\{\mathbf{r}_i^{N_2}\}$ is defined as
\begin{equation} \label{eq:density 2}
  \rho_i^{(2)}[\{C,\alpha\}] = (1 +
  \alpha^NG_{ii})^{-3/2}\exp\left[
    -\frac{3\alpha^N}{2a^2}\frac{(\mathbf{s}_i - 
      \mathbf{r}_i^{N_2})^2}{1 +
      \alpha^NG_{ii}}\right].
\end{equation} 
The structural similarity relative to the native structures given by
$\{\rho^{(1)}\}$ and $\{\rho^{(2)}\}$specify local order parameters
suitable to describing conformational transitions between metastable states in
proteins.

To investigate the detailed main-chain dynamics controlling the
structural change in CaM, we characterize the relative similarity to the
closed structure along the transition route through the normalized measure
\begin{equation} \label{eq:norm density 1}
  \overline{\rho_i}^{(1)}(\alpha_0) = 
  \frac{\rho_i^{(1)}(\alpha_0) - \rho_i^{(1)}(0)}
       {\rho_i^{(1)}(1) - \rho_i^{(1)}(0)},
\end{equation}
where $\rho_i^{(1)}(\alpha_0)$ is the monomer density of the $i^{\mathrm{th}}$ residue
with respect to the closed conformation (Eq.~\ref{eq:density 1}).
Similarly, we represent the relative structural similarity to the 
open conformation as
\begin{equation} \label{eq:norm density 2}
  \overline{\rho_i}^{(2)}(\alpha_0) 
  = \frac{\rho_i^{(2)}(\alpha_0) - \rho_i^{(2)}(1)}
           {\rho_i^{(2)}(0) - \rho_i^{(2)}(1)},
\end{equation}
where $\rho_i^{(1)}(\alpha_0)$ is the monomer density of the $i^{\mathrm{th}}$ residue
with respect to the open conformation (Eq.~\ref{eq:density 2}).
In the open state, $\overline{\rho_i}^{(1)}(0)=0$ and 
$\overline{\rho_i}^{(2)}(0)=1$, while in the closed state
$\overline{\rho_i}^{(1)}(1)=1$ and 
$\overline{\rho_i}^{(2)}(1)=0$.
To represent the structural changes more clearly, it is convenient
to consider the difference,
\begin{equation}\label{eq:natdens_diff}
\Delta\overline{\rho_i}(\alpha_0) =
\overline{\rho_i}^{(1)}(\alpha_0) - \overline{\rho_i}^{(2)}(\alpha_0) 
\end{equation}
for each residue. This difference shifts the relative degree of localization
to be between $\Delta\overline{\rho_i}(1) = 1$ and 
$\Delta\overline{\rho_i}(0) =-1$ corresponding to the open and closed conformations,
respectively.

\section*{RESULTS}
\subsection*{Conformational Flexibility and Calcium Binding}

\begin{figure}
   \begin{center}
      \includegraphics[width=6.0in]{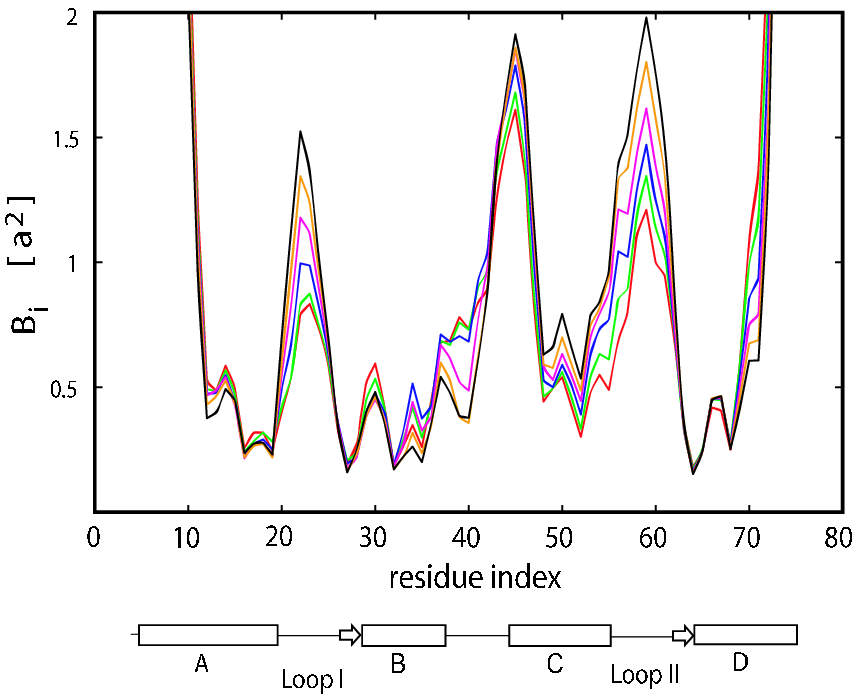}
      \caption{}
      \label{fig:Gii}
   \end{center}
\end{figure}

\begin{figure}
   \begin{center}
      \includegraphics[width=4.5in]{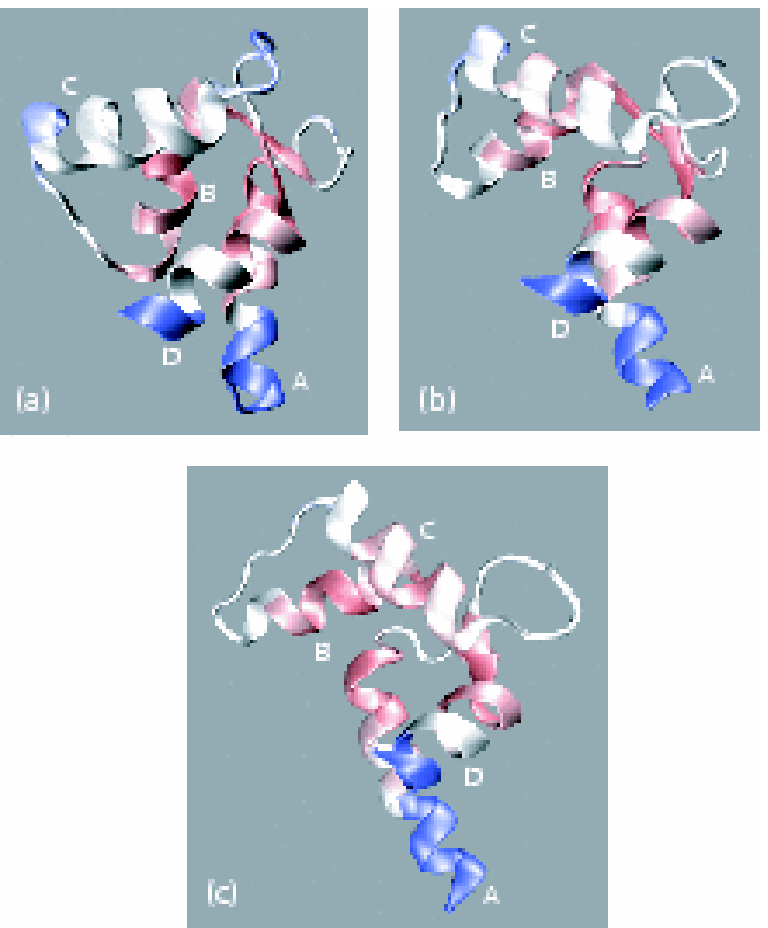}
      \caption{}
      \label{fig:B_3D}
   \end{center}
\end{figure}

The local mean square fluctuations of $\alpha$-carbon positions
(related to the temperature factors from X-ray crystallography) are a
natural set of order parameters for the reference Hamiltonian $H_0$ in
our model. This parameter, $B_i =
\langle\delta\mathbf{r}_i^2\rangle_0$, contains information about the
degree of structural order and conformational flexibility of each
residue. In Fig.~\ref{fig:Gii} we have plotted $B_i$ versus sequence
number at different values of $\alpha_0$, the parameter that controls
the uniform interpolation between the open structure ($\alpha_0 = 0$)
and the closed structure ($\alpha_0 = 1$). Fig.~\ref{fig:B_3D} shows
the corresponding 3D structures of nCaM domain with the residues
colored according to $B_i$. Aside from the very flexible ends of two
terminal helices A and D, the Ca$^{2+}$-binding
loops and the helix linker possess the highest flexibility. The calculated
fluctuations from our model exhibit very good qualitative
agreement with X-ray temperature factors~\cite{grabarek:05} and
simulation results~\cite{zuckerman:04,likic:gooley:03} of CaM.

\textbf{Binding loops.}~ Each EF-hand in CaM coordinates Ca$^{2+}$ through 
a 12-residue loop:
Asp20-Glu31 in loop I and Asp56-Glu67 in loop II.  The C-terminal ends
of the loops contain a short $\beta$-sheet (residues 26-28 in loop I
and residues 62-64 in loop II) adjacent the last three residues that
are part of the exiting helices B and D, respectively. 

As shown in Fig.~\ref{fig:Gii}, the loops remain relatively flexible
even in the open conformation. The highest flexibility is near the two 
Glycines in position 4 of the Ca$^{2+}$-binding loops I (Gly23) and 
II (Gly59). This invariable Gly residue provides a sharp turn required 
for the proper geometry of the Ca$^{2+}$-binding 
sites.\cite{strynadka:james:89,likic:gooley:03} The linker between
helices B and C is also very mobile, with the highest flexibility near
residue Glu45. Taken together, the mobility of the loops
and B/C linker indicates that the domain opening depends entirely on a
set of inherent dynamics, or ``intrinsic plasticity'', of
CaM.\cite{chou:bax:01}

A closer look at the fluctuations of the Ca$^{2+}$-binding loops reveals
that the N-terminal part of each loop is more flexible than the
C-terminal part. This agrees with NMR data characterizing the
flexibility of the N-terminal and C-terminal part of loop III and IV of the 
C-terminal domain.\cite{malmendal:akke:99,evenas:akke:99} In the transition 
route (from closed $\rightarrow$ open), the N-terminal ends of the loops stiffen 
gradually. On the other hand, in the C-terminal part of the loops the short
$\beta$-sheet structure (residues 26-28 in loop I and 62-64 in loop II) 
remain rigid (see Fig.~\ref{fig:Gii} and~\ref{fig:B_3D}). 
Also the last three residues of the loops (residues 29-31 in loop I/helix B 
and residues 65-67 in loop II/helix D) remain relatively rigid, stabilized by the 
exiting helices B and D respectively.\cite{strynadka:james:89}

This immobile region, the EF-hand $\beta$-scaffold, is central to a recent proposed 
mechanism for CaM~\cite{grabarek:05} and other EF-hand domains.\cite{grabarek:06}
Fig.~\ref{fig:Gii} shows that residues Thr26 and Ile27 (in $\beta$-sheet of loop I) and 
Thr62 and Ile63 (in $\beta$-sheet of loop II) remain 
very rigid during the domain opening. 

It is also interesting to compare the relative flexibility of binding
loop I and II. It is clear that binding loop II is more flexible than
loop I in the both conformations (see Fig.~\ref{fig:Gii} and
~\ref{fig:B_3D}(a)). In particular, the connection between helix A and
the binding loop I is much more rigid than the connection between
helix C and the binding loop II.  This large difference in flexibility
suggests that binding loop II of nCaM is more dominate in the
mechanism for the structural transition. A similar mechanism in
C-terminal CaM domain was also observed from NMR studies, where the
Ca$^{2+}$-dependent exchange contribution is dominated by binding loop
IV with lower $S^{2}$ (higher flexibility) than loop
III.\cite{malmendal:akke:99}

\textbf{Helices B and C and the B/C linker.}~ Fig.~\ref{fig:Gii} and 
Fig.~\ref{fig:B_3D} also shows that the bottom part of helix C (close
to B/C helix linker) is very flexible in apo nCaM. Upon opening, the
flexibility of helix C decreases significantly. [See the change in
color from blue to white (Fig.~\ref{fig:B_3D}(a)-(c)) at the bottom
part (close to B/C helix-linker) of helix C and from white to red at
the middle part of helix C.] In contrast, the top part of helix B
(close to binding loop I; residues 29--31) becomes more flexible than
the bottom part of helix B (close to B/C helix-linker; residues 32--37)
during closed to open transition (see Fig.~\ref{fig:Gii}). We also
note that residues 37--42 of the B/C helix-linker shows significant
increase in flexibility during opening of the domain.  This change in
flexibility of the B/C helix-linker helps facilitate the concerted
reorientation of helices B and C during the closed $\rightarrow$ open
transition.  Similar behavior was also observed in molecular dynamics
simulation of CaM~\cite{wriggers:schulten:98} for this six-residue
(residues 37--42) segment.

\subsection*{Conformational Change and Transition Mechanism}

The results discussed in the previous section gives a picture of the
closed to open transition with good overall agreement with experiment
and simulation results on an isolated apo-CaM domain. Nevertheless,
the analysis has focused primarily on the difference in the magnitude
of fluctuations of the two meta-stable states. We now turn our attention
to the predicted transition mechanism and qualitative nature of
structural changes along the transition route. Such a description includes:
along the transition route from closed to open, what structural
changes are predicted to occur early/late, and
which are predicted to happen gradually/cooperatively. While such details
have yet to be revealed directly through measurement, in principle, site-directed
mutagenesis experiments can be used to identify kinetically important
structural regions of nCaM.

\begin{figure}
   \begin{center}
      \includegraphics[width=6.0in]{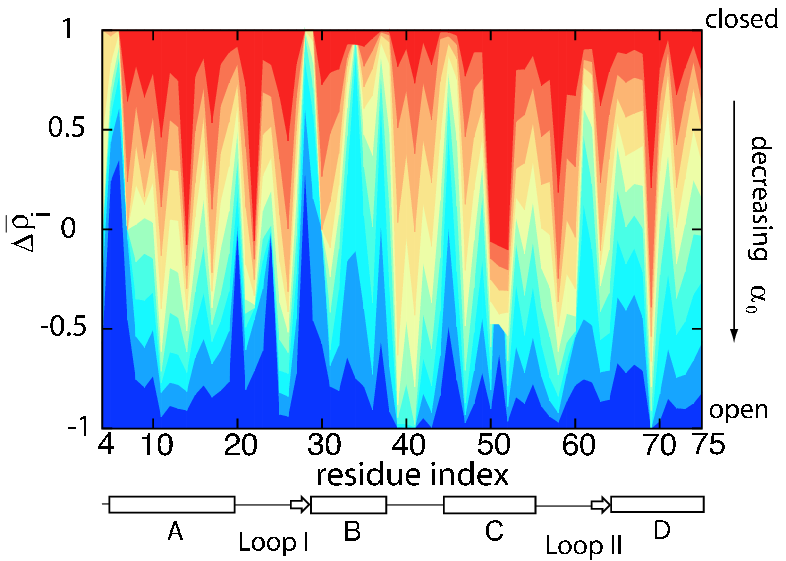}
      \caption{}
      \label{fig:drhon}
   \end{center}
\end{figure}

\begin{figure}
   \begin{center}
      \includegraphics[width=4.5in]{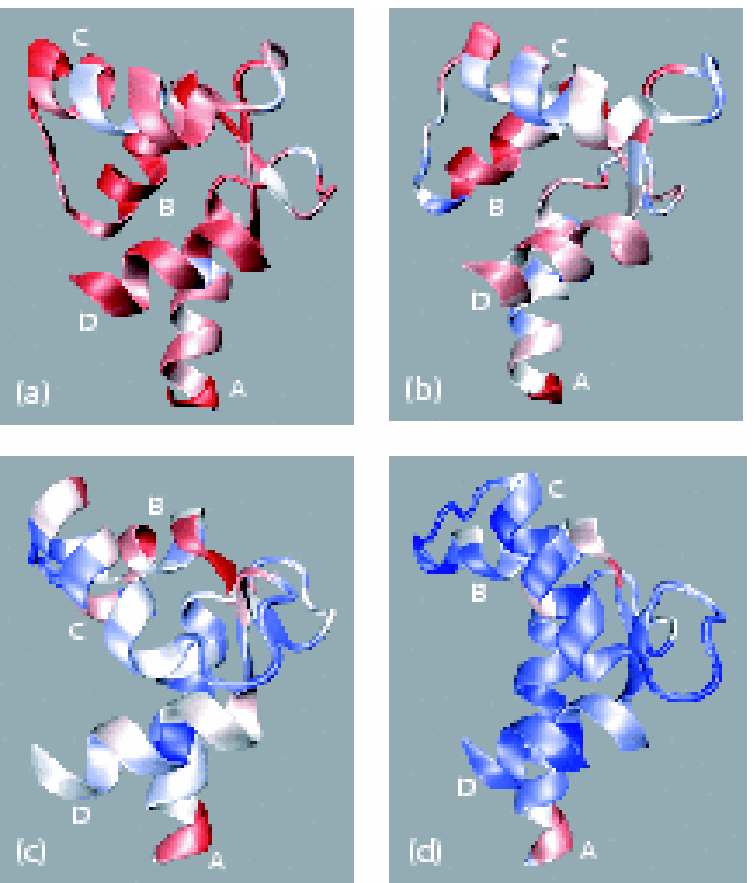}
      \caption{}
      \label{fig:drhon_3D}
   \end{center}
\end{figure}

To clarify the transtition route, we introduce a structural order parameter that measures
the similarity to the open or closed state, $\Delta\overline{\rho_i}$
given in Eq.~\ref{eq:natdens_diff}. This order parameter is defined
so that $\Delta\overline{\rho_i}=1$ corresponds to the closed
conformation and $\Delta\overline{\rho_i}=-1$ corresponds to the open
conformation of nCaM domain.  Fig.~\ref{fig:drhon} illustrates the
conformational transition in nCaM domain in terms of
$\Delta\overline{\rho_i}$ for each residue. An alternative
representation of the same data is shown in Fig.~\ref{fig:drhon_3D};
here, the value of $\Delta\overline{\rho_i}$ is represented as colors
ranging from red ($\Delta\overline{\rho_i} = -1$) to white
($\Delta\overline{\rho_i}$ = 0) to blue ($\Delta\overline{\rho_i} =
-1$) superimposed on the interpolated structure for selected
values of $\alpha_0$.

We first notice  that an early
transition in the binding loops and in the central region of helix C
evident in Fig.~\ref{fig:drhon}. [See also the gradual change in color from blue to red in the
structures of Fig.~\ref{fig:drhon_3D}(a)-(d).]
We also note the concerted structural change
of parts of helices B and C and
flexible B/C helix-linker (residues 31--49).
In particular, the flexible B/C
helix-linker (residues 38-44) in Fig.~\ref{fig:drhon} exhibits a cooperative
transition. Residue Gln41 which located in this linker region is
highly mobile according to NMR data.\cite{kuboniwa:bax:95,zhang:ikura:95} 
The change in color from red to blue
in the B/C helix linker in Fig.~\ref{fig:drhon_3D}(a) and (b) indicates
that the structural transition of the 
N-terminal part (close to helix B) of this linker occurs earlier its 
C-terminal part (close to helix C). 

Fig.~\ref{fig:drhon} and Fig.~\ref{fig:drhon_3D} also show a delayed
initiation of structural change in residues 4--7 of helix A, residues
27--30 of binding loop I and N-terminal part of helix B.
Specifically, the residues near the top part of helix B (close to
binding loop I) and in binding loop I, have very little structural
change at the beginning of domain opening, with a sharp, cooperative
transition near the end. [See the relatively slow color change (from
red to blue) in this part of helix B and binding loop I in
Fig.~\ref{fig:drhon_3D}(a)-(d).] Although, the middle part of helix C
(residues 50--52) has some limited structural change early in the
transition, it remains quite immobile after that. [See
Fig.~\ref{fig:drhon} and the early color change from red to blue in
Fig.~\ref{fig:drhon_3D}.]

\begin{figure}
   \begin{center}
      \includegraphics[width=4.5in]{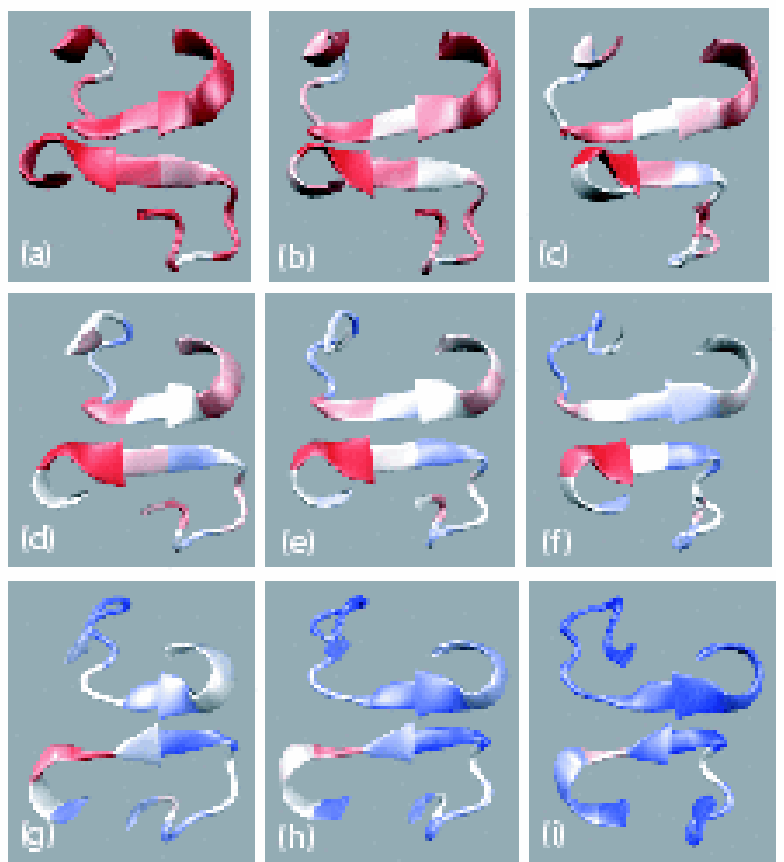}
      \caption{}
      \label{fig:loops_3D}
   \end{center}
\end{figure}

\textbf{Binding loops I and II.} Because of the central importance
of the interactions between the binding loops in the recently proposed
two-step Ca$^{2+}$-binding mechanism, this EF$\beta$-scaffold region
is highlighted in Fig.~\ref{fig:loops_3D}. In the first step of this
binding mechanism, the Ca$^{2+}$ is immobilized by the structural
rigidity in the plane of $\beta$-sheet and the ligands from N-terminal
part of the binding loops. In the second step, the backbone torsional
flexibility of the EF$\beta$-scaffold enables repositioning of the
C-terminal part of the binding loop together with the exiting helix
(helix B in loop I and helix D in loop II).\cite{grabarek:06} Since
the Ca$^{2+}$ ions are not included in our model and we can not
characterize backbone torsional flexibility of the EF$\beta$-scaffold,
our analysis is independent of that developed in Ref.~\onlinecite{grabarek:05,grabarek:06}.
The closed to open conformational
transition of each binding loop is quite different in
Fig.~\ref{fig:loops_3D}. We predict that the structural
changes in binding loop II occur before binding loop I upon domain
opening (see the relatively slow color change from red to blue in
binding loop I than loop II in Fig.~\ref{fig:loops_3D}). 
Since the
flexibility of binding loop II is also greater, this suggests that
during Ca$^{2+}$-binding process the loop II is more dominates
the overall conformational change between the
closed and open state. This agrees with results based on the all atom
molecular dynamics simulations of nCaM discussed by Vigil et al..\cite{vigil:garcia:01}

\begin{figure}
   \begin{center}
      \includegraphics[width=5.0in]{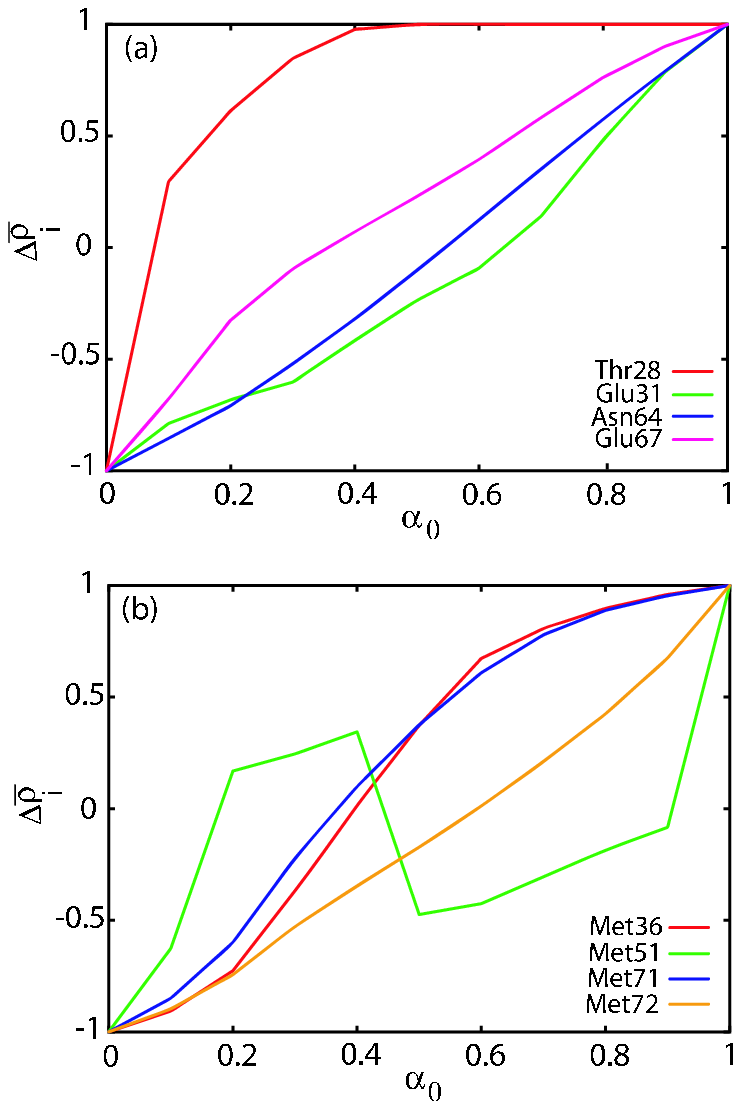}
      \caption{}
      \label{fig:residues}
   \end{center}
\end{figure}

Fig.~\ref{fig:loops_3D} also shows that the N-terminal ends of the
loops have relatively an early transition compared to the C-terminal
ends. Furthermore, the conformation change of the C-terminal end of
binding loop I is more cooperative, presumably relying on the earlier
structural change in binding loop II. Specifically, the closed state
structure residue in position 9 (Thr28) of the loop I is very stable
as shown in Fig.~\ref{fig:residues}(a). This is due to a hydrogen bonding
between Thr28 and Glu31. Fig.~\ref{fig:residues}(a) also suggests that
the structural change of Glu31 occurs before Thr28 upon domain
opening, and proceeds through the transition much more
gradually. Similar hydrogen bonding is also present between Asn64 and
Glu67 in binding loop II. Nevertheless, compared to the corresponding
residues in loop I, the structural change of these two residues is
quite gradual [see Fig.~\ref{fig:residues}(a)]. Nevertheless, Asn64 does seem
to have a somewhat sharper transition than Glu67. Finally, residues
Gly61 and Thr62 in binding loop II exhibit little structural
change in Fig.~\ref{fig:loops_3D} as the domain begins to open.

\textbf{Methionine residues.}  The large hydrophobic binding surfaces
that open in both domains of CaM are especially rich in Methionine
residues, with four Methionines in each domain occupying nearly 46\%
of the total hydrophobic surface area.\cite{nelson:chazin:98} These
side chains as well as other aliphatic residues, such as Valine,
Isoleucine and Leucine, which make up the rest of the hydrophobic
binding surface are highly dynamic in
solution.\cite{siivari:vogel:95} The flexibility of the residues
composing hydrophobic binding surface for target peptides explains
CaM's high degree of binding promiscuity. Here we consider the
main-chain flexibility. The four Methionine residues in nCaM are
situated in position 36, 51, 71 and 72.  The closed to open structural
transition of residues Met36 and Met71 are similar and relatively
sharp compared to residue Met72 which is quite gradual as shown in
Fig.~\ref{fig:residues}(b). This suggests that residues Met36 and
Met71 remains relatively buried in the beginning of the domain
opening. Curiously, from Fig.~\ref{fig:residues}(b) residue Met51 in
the middle part of helix C at $\alpha_0=0.5$, shows sudden increase in
$\Delta\overline{\rho_i}$ during closed to open conformational change.

\subsection*{Conformational Transition Rate and Order Parameter}

The one dimensional free energy profile parameterized by the
interpolation parameter $\alpha_0$ is shown in Fig.~\ref{fig:f_Q}.
The minimum corresponding to the open state is very shallow and unstable
compared to the closed state.  Combined molecular
dynamics simulations and small angle X-ray scattering studies on apo
nCaM and Ca$^{2+}$-bound nCaM by Vigil et al.~\cite{vigil:garcia:01}
have also shown that in aqueous solution the closed state dominates
the population.  The equilibrium populations for the closed and open
state from our model are found to be 94\% and 6\% respectively. For
comparison, the NMR measurement of apo cCaM indicate a minor
population of 5--10\%.\cite{malmendal:akke:99} These results suggest
that on average, the residues in the hydrophobic surface of CaM are well
protected from solvent.

The maximum of the free energy occurs
quite close to the open state at $\alpha_0 \sim 0.2$, though the
barrier is very broad in terms of this reaction coordinate. 
We also consider
the free energy of the global structural parameter $\Delta Q =
Q_1 - Q_2 = \sum\Delta\overline{\rho_i}/N$ where
$\Delta\overline{\rho_i}$ is given in in Eq.~\ref{eq:natdens_diff}.  
Fig.~\ref{fig:f_Q} shows that 
$\Delta Q$ is also a reasonable reaction coordinate for the
transition. The barrier broadens somewhat, with
the maximum free energy occurring around $\Delta Q = -0.25$.  In terms of the global
structure, this 
roughly corresponds to 60\%--75\% of nCaM being
similar to open state configuration in the transition state ensemble.

Even though the open state minimum is not well isolated,
we estimate the conformational
transition rate from closed to open using the Arrhenius form,
$k=k_0e^{-\Delta{F^\dagger}/k_{B}T}$ where $\Delta{F^\dagger}$ is the
free energy difference between the closed conformation and
transition-state ensemble. Assuming the prefactor $k_0=1\mu$s$^{-1}$
gives the estimate $k=40,000$s$^{-1}$. This value is in reasonable
agreement with the transition rate estimate of 
$k=20,000$s$^{-1}$ based on NMR exchange rate data of cCaM.\cite{malmendal:akke:99}

\section*{DISCUSSION}

The primary motivation for the work presented in this paper
is to understand protein functions that involve large scale (main-chain) dynamics and
flexibility. Proteins with relatively large conformational freedom
include those in which folding and binding are coupled.\cite{shoemaker:wolynes:00},
as well as
hinge bending motions~\cite{sinha:nussinov:01} or proteins with high
plasticity such as ion binding
sites,\cite{bentley:rety:00} and proteins with
allosteric transitions.\cite{lundstrom:akke:05} While not nearly as
developed as the Energy Landscape Theory of protein
folding,\cite{bryngelson:wolynes:95} a general
thermodynamic framework for the Energy Landscape Theory of
protein-protein
binding,\cite{verkhivker:rose:02,wang:wang:06}
large conformational transitions,\cite{sinha:nussinov:01} and the
coupling between folding and binding~\cite{papoian:wolynes:03} is
beginning to emerge. Aside from some noted
exceptions,\cite{verkhivker:rose:03,miyashita:wolynes:03,levy:onuchic:04,yang:onuchic:04,levy:wolynes:05,best:hummer:05}
relatively little theoretical work has focused on detailed analysis of
transition mechanisms of flexible proteins in terms of specific
ensembles of kinetic pathways. The dynamics of conformational
transitions between well-defined conformational basins are generally
controlled by relatively low probability partially ordered
ensembles. The main challenge is to describe the transition state
ensembles at the residue level giving a site-specific description of
the transition mechanism.

Modern NMR relaxation experiments have provided a wealth of data about
internal dynamics and conformational sub-states quantitatively on fast
(nanosecond) and slow (micro- to millisecond)
timescales.\cite{ishima:torchia:00} Such studies 
are very useful in identifying residues with high flexibility upon
target binding, not only through movements of surface loops and side
chains, but also by global motions of the core
structure.\cite{huang:montelione:05} These experiments, however,
provide only a few local structural changes and have not been able to
capture the molecular details necessary to fully understand the
mechanism of conformational transitions. Whereas atomistic simulations
can potentially bridge the gap on time
scale up to microsecond, this timescale falls orders of magnitude short for slow
protein dynamics (millisecond to second). Also, the use of atomistic
approaches becomes computationally inefficient with the increased size
of a system.

To overcome the problems associated with all-atom simulations, many
studies has demonstrated the use of coarse-grained protein models with
simplified representations, such as, only $\alpha$-carbons as point
masses and simplified energy functions.\cite{tozzini:05} Such models require
much less computational cost making them practical to describe the
conformational transitions of even 
large proteins.\cite{kim:chirikjian:02} Analyzing the
fluctuations about a single minimum has been surprisingly successful
in identifying relevant cooperative motions in a wide range of
proteins. The commonly used Tirion
potential~\cite{tirion:96,bahar:erman:98} (which can be viewed as a
harmonic Go-model) gives a simple one parameter model in which the
relevant motions for the transition is identified as one of many low
frequency normal modes.\cite{tama:brooks:02} While
this approach can provide considerable insight, it offers a limited
description of the transition because it is based only on the
fluctuations about one structure. The Tirion potential has recently
been extended to include two conformations in which the contact map
defining the potential and normal modes is updated as the protein is
moved along a known reaction
coordinate.\cite{miyashita:wolynes:03,miyashita:wolynes:04} Local
unfolding and flexibility is accommodated by relieving regions of high
stress, ``cracking'', which modifies the contact map. Coarse-grained 
simulations in which the potential interpolates between two
folded-state biased contact maps have also been introduced
recently.\cite{zuckerman:04,best:hummer:05,maragakis:karplus:05,okazaki:wolynes:06}
For example, in the plastic network model of Margakis 
and Karplus~\cite{maragakis:karplus:05} the individual basins are approximated 
by the Tirion potential and are then smoothly connected by a secular equation 
formulation. A similar interpolation was considered by Okazaki et al.~\cite{okazaki:wolynes:06} 
Alternatively, Best et al. developed a two-state 
approximation~\cite{best:hummer:05} analogous to Eq.~\ref{eq:contact}.
These advances are similar in spirt to our approach, albeit with distinct approximations
for the basic description of partially ordered ensembles.

\section*{CONCLUSION} 

In this paper, we study the intrinsic flexibility and structural change in the 
N-terminal domain of CaM (nCaM) during open to close transition.
The predicted transition route from our model
gives a detailed picture of the
interplay between structural transition, conformational flexibility
and function of N-terminal calmodulin (nCaM) domain.
The results from our model are largely consistent with the important 
role that the immobile EF$\beta$-scaffold region plays in the transition 
mechanism. Dissection of the transition route of this region further 
suggests that it is the early structural change of loop II that drives 
the cooperative completion of the interactions between the loops in 
the open structure. 

The strong qualitative agreement with available experimental
measurements of flexibility is an encouraging validation of the model.
Recently, the folding dynamics of zinc-metallated protein (azurin) was
studied using a similar variational model and compared with
experiments for the detail coordination reaction coupled with the
entatic state.\cite{zong:wolynes:07} A similar future study of 
detail coordination reaction for the complete description of conformational 
change stabilized by ion binding in CaM seems very promising. Ultimately, 
we wish to extend this model to investigate the binding mechanism and kinetic 
paths of several peptides to Ca$^{2+}$-loaded CaM. Since large conformational 
changes coupled to binding depends fundamentally on the fluctuations of partially
folded conformations,\cite{shoemaker:wolynes:00} this polymer based variational 
formalism can accommodate coupled folding and binding very naturally.

\section*{Acknowledgments}

We thank Zenon Grabarek for helpful suggestions and critically reading the manuscript.
This work was supported in part by grant awarded by the Ohio Board of 
Regents Research Challenge program.

\newpage

\textbf{REFERENCES}


\begin{thebibliography}{53}
\expandafter\ifx\csname natexlab\endcsname\relax\def\natexlab#1{#1}\fi
\expandafter\ifx\csname bibnamefont\endcsname\relax
  \def\bibnamefont#1{#1}\fi
\expandafter\ifx\csname bibfnamefont\endcsname\relax
  \def\bibfnamefont#1{#1}\fi
\expandafter\ifx\csname citenamefont\endcsname\relax
  \def\citenamefont#1{#1}\fi
\expandafter\ifx\csname url\endcsname\relax
  \def\url#1{\texttt{#1}}\fi
\expandafter\ifx\csname urlprefix\endcsname\relax\def\urlprefix{URL }\fi
\providecommand{\bibinfo}[2]{#2}
\providecommand{\eprint}[2][]{\url{#2}}

\bibitem[{\citenamefont{Frauenfelder et~al.}(1991)\citenamefont{Frauenfelder,
  Sligar, and Wolynes}}]{frauenfelder:wolynes:91}
\bibinfo{author}{\bibfnamefont{H.}~\bibnamefont{Frauenfelder}},
  \bibinfo{author}{\bibfnamefont{S.~G.} \bibnamefont{Sligar}},
  \bibnamefont{and} \bibinfo{author}{\bibfnamefont{P.~G.}
  \bibnamefont{Wolynes}}, \bibinfo{journal}{Science}
  \textbf{\bibinfo{volume}{254}}, \bibinfo{pages}{1598} (\bibinfo{year}{1991}).

\bibitem[{\citenamefont{Gerstein et~al.}(1994)\citenamefont{Gerstein, Lesk, and
  Chothia}}]{gerstein:chothia:94}
\bibinfo{author}{\bibfnamefont{M.}~\bibnamefont{Gerstein}},
  \bibinfo{author}{\bibfnamefont{A.~M.} \bibnamefont{Lesk}}, \bibnamefont{and}
  \bibinfo{author}{\bibfnamefont{C.}~\bibnamefont{Chothia}},
  \bibinfo{journal}{Biochemistry} \textbf{\bibinfo{volume}{33}},
  \bibinfo{pages}{6739} (\bibinfo{year}{1994}).

\bibitem[{\citenamefont{Nelson and Chazin}(1998)}]{nelson:chazin:98}
\bibinfo{author}{\bibfnamefont{M.~R.} \bibnamefont{Nelson}} \bibnamefont{and}
  \bibinfo{author}{\bibfnamefont{W.~J.} \bibnamefont{Chazin}},
  \bibinfo{journal}{Protein Sci.} \textbf{\bibinfo{volume}{7}},
  \bibinfo{pages}{270} (\bibinfo{year}{1998}).

\bibitem[{\citenamefont{Brokx et~al.}(2001)\citenamefont{Brokx, Lopez, Vogel,
  and Makhatadze}}]{brokx:makhatadze:01}
\bibinfo{author}{\bibfnamefont{R.~D.} \bibnamefont{Brokx}},
  \bibinfo{author}{\bibfnamefont{M.~M.} \bibnamefont{Lopez}},
  \bibinfo{author}{\bibfnamefont{H.~J.} \bibnamefont{Vogel}}, \bibnamefont{and}
  \bibinfo{author}{\bibfnamefont{G.}~\bibnamefont{Makhatadze}},
  \bibinfo{journal}{J. of Biol. Chem.} \textbf{\bibinfo{volume}{276}},
  \bibinfo{pages}{14083} (\bibinfo{year}{2001}).

\bibitem[{\citenamefont{Chou et~al.}(01)\citenamefont{Chou, Li, Klee, and
  Bax}}]{chou:bax:01}
\bibinfo{author}{\bibfnamefont{J.~J.} \bibnamefont{Chou}},
  \bibinfo{author}{\bibfnamefont{S.}~\bibnamefont{Li}},
  \bibinfo{author}{\bibfnamefont{C.~B.} \bibnamefont{Klee}}, \bibnamefont{and}
  \bibinfo{author}{\bibfnamefont{A.}~\bibnamefont{Bax}},
  \bibinfo{journal}{Nat.\ Struct.\ Biol.} \textbf{\bibinfo{volume}{8}},
  \bibinfo{pages}{990} (\bibinfo{year}{01}).

\bibitem[{\citenamefont{Tsalkova and Privalov}(1985)}]{tsalkova:privalov:85}
\bibinfo{author}{\bibfnamefont{T.~N.} \bibnamefont{Tsalkova}} \bibnamefont{and}
  \bibinfo{author}{\bibfnamefont{P.~L.} \bibnamefont{Privalov}},
  \bibinfo{journal}{J.\ Mol.\ Biol.} \textbf{\bibinfo{volume}{181}},
  \bibinfo{pages}{533} (\bibinfo{year}{1985}).

\bibitem[{\citenamefont{Linse et~al.}(1991)\citenamefont{Linse, Helmersson, and
  Forsen}}]{linse:forsen:91}
\bibinfo{author}{\bibfnamefont{S.}~\bibnamefont{Linse}},
  \bibinfo{author}{\bibfnamefont{A.}~\bibnamefont{Helmersson}},
  \bibnamefont{and} \bibinfo{author}{\bibfnamefont{S.}~\bibnamefont{Forsen}},
  \bibinfo{journal}{J.\ Biol.\ Chem.} \textbf{\bibinfo{volume}{266}},
  \bibinfo{pages}{8050} (\bibinfo{year}{1991}).

\bibitem[{\citenamefont{Baber et~al.}(2001)\citenamefont{Baber, Szabo, and
  Tjandra}}]{baber:tjandra:01}
\bibinfo{author}{\bibfnamefont{J.~L.} \bibnamefont{Baber}},
  \bibinfo{author}{\bibfnamefont{A.}~\bibnamefont{Szabo}}, \bibnamefont{and}
  \bibinfo{author}{\bibfnamefont{N.}~\bibnamefont{Tjandra}},
  \bibinfo{journal}{J.\ Am.\ Chem.\ Soc.} \textbf{\bibinfo{volume}{123}},
  \bibinfo{pages}{3953} (\bibinfo{year}{2001}).

\bibitem[{\citenamefont{Malmendal et~al.}(1999)\citenamefont{Malmendal, Evanas,
  Forsen, and Akke}}]{malmendal:akke:99}
\bibinfo{author}{\bibfnamefont{A.}~\bibnamefont{Malmendal}},
  \bibinfo{author}{\bibfnamefont{J.}~\bibnamefont{Evanas}},
  \bibinfo{author}{\bibfnamefont{S.}~\bibnamefont{Forsen}}, \bibnamefont{and}
  \bibinfo{author}{\bibfnamefont{M.}~\bibnamefont{Akke}}, \bibinfo{journal}{J.\
  Mol.\ Biol.} \textbf{\bibinfo{volume}{293}}, \bibinfo{pages}{883}
  (\bibinfo{year}{1999}).

\bibitem[{\citenamefont{Ishima and Torchia}(2000)}]{ishima:torchia:00}
\bibinfo{author}{\bibfnamefont{R.}~\bibnamefont{Ishima}} \bibnamefont{and}
  \bibinfo{author}{\bibfnamefont{D.~A.} \bibnamefont{Torchia}},
  \bibinfo{journal}{Nat.\ Struct.\ Biol.} \textbf{\bibinfo{volume}{7}},
  \bibinfo{pages}{740} (\bibinfo{year}{2000}).

\bibitem[{\citenamefont{Evenas et~al.}(1999)\citenamefont{Evenas, Forsen,
  Malmendal, and Akke}}]{evenas:akke:99}
\bibinfo{author}{\bibfnamefont{J.}~\bibnamefont{Evenas}},
  \bibinfo{author}{\bibfnamefont{S.}~\bibnamefont{Forsen}},
  \bibinfo{author}{\bibfnamefont{A.}~\bibnamefont{Malmendal}},
  \bibnamefont{and} \bibinfo{author}{\bibfnamefont{M.}~\bibnamefont{Akke}},
  \bibinfo{journal}{J.\ Mol.\ Biol.} \textbf{\bibinfo{volume}{289}},
  \bibinfo{pages}{603} (\bibinfo{year}{1999}).

\bibitem[{\citenamefont{Grabarek}(2005)}]{grabarek:05}
\bibinfo{author}{\bibfnamefont{Z.}~\bibnamefont{Grabarek}},
  \bibinfo{journal}{J.\ Mol.\ Biol.} \textbf{\bibinfo{volume}{346}},
  \bibinfo{pages}{1351} (\bibinfo{year}{2005}).

\bibitem[{\citenamefont{Grabarek}(2006)}]{grabarek:06}
\bibinfo{author}{\bibfnamefont{Z.}~\bibnamefont{Grabarek}},
  \bibinfo{journal}{J.\ Mol.\ Biol.} \textbf{\bibinfo{volume}{359}},
  \bibinfo{pages}{509} (\bibinfo{year}{2006}).

\bibitem[{\citenamefont{Kuboniwa et~al.}(1995)\citenamefont{Kuboniwa, Tjandra,
  Grzesiek, Ren, Klee, and Bax}}]{kuboniwa:bax:95}
\bibinfo{author}{\bibfnamefont{H.}~\bibnamefont{Kuboniwa}},
  \bibinfo{author}{\bibfnamefont{N.}~\bibnamefont{Tjandra}},
  \bibinfo{author}{\bibfnamefont{S.}~\bibnamefont{Grzesiek}},
  \bibinfo{author}{\bibfnamefont{H.}~\bibnamefont{Ren}},
  \bibinfo{author}{\bibfnamefont{C.~B.} \bibnamefont{Klee}}, \bibnamefont{and}
  \bibinfo{author}{\bibfnamefont{A.}~\bibnamefont{Bax}},
  \bibinfo{journal}{Nat.\ Struct.\ Biol.} \textbf{\bibinfo{volume}{2}},
  \bibinfo{pages}{768} (\bibinfo{year}{1995}).

\bibitem[{\citenamefont{Zhang et~al.}(1995)\citenamefont{Zhang, Tanaka, and
  Ikura}}]{zhang:ikura:95}
\bibinfo{author}{\bibfnamefont{M.}~\bibnamefont{Zhang}},
  \bibinfo{author}{\bibfnamefont{T.}~\bibnamefont{Tanaka}}, \bibnamefont{and}
  \bibinfo{author}{\bibfnamefont{M.}~\bibnamefont{Ikura}},
  \bibinfo{journal}{Nat.\ Struct.\ Biol.} \textbf{\bibinfo{volume}{2}},
  \bibinfo{pages}{758} (\bibinfo{year}{1995}).

\bibitem[{\citenamefont{Portman et~al.}(1998)\citenamefont{Portman, Takada, and
  Wolynes}}]{portman:wolynes:98}
\bibinfo{author}{\bibfnamefont{J.~J.} \bibnamefont{Portman}},
  \bibinfo{author}{\bibfnamefont{S.}~\bibnamefont{Takada}}, \bibnamefont{and}
  \bibinfo{author}{\bibfnamefont{P.~G.} \bibnamefont{Wolynes}},
  \bibinfo{journal}{Phys.\ Rev.\ Lett.} \textbf{\bibinfo{volume}{81}},
  \bibinfo{pages}{5237} (\bibinfo{year}{1998}).

\bibitem[{\citenamefont{Portman
  et~al.}(2001{\natexlab{a}})\citenamefont{Portman, Takada, and
  Wolynes}}]{portman:wolynes:01a}
\bibinfo{author}{\bibfnamefont{J.~J.} \bibnamefont{Portman}},
  \bibinfo{author}{\bibfnamefont{S.}~\bibnamefont{Takada}}, \bibnamefont{and}
  \bibinfo{author}{\bibfnamefont{P.~G.} \bibnamefont{Wolynes}},
  \bibinfo{journal}{J.\ Chem.\ Phys.} \textbf{\bibinfo{volume}{114}},
  \bibinfo{pages}{5069} (\bibinfo{year}{2001}{\natexlab{a}}).

\bibitem[{\citenamefont{Portman
  et~al.}(2001{\natexlab{b}})\citenamefont{Portman, Takada, and
  Wolynes}}]{portman:wolynes:01b}
\bibinfo{author}{\bibfnamefont{J.~J.} \bibnamefont{Portman}},
  \bibinfo{author}{\bibfnamefont{S.}~\bibnamefont{Takada}}, \bibnamefont{and}
  \bibinfo{author}{\bibfnamefont{P.~G.} \bibnamefont{Wolynes}},
  \bibinfo{journal}{J.\ Chem.\ Phys.} \textbf{\bibinfo{volume}{114}},
  \bibinfo{pages}{5082} (\bibinfo{year}{2001}{\natexlab{b}}).

\bibitem[{\citenamefont{Wriggers et~al.}(1998)\citenamefont{Wriggers, Mehler,
  Pitici, Weinstein, and Schulten}}]{wriggers:schulten:98}
\bibinfo{author}{\bibfnamefont{W.}~\bibnamefont{Wriggers}},
  \bibinfo{author}{\bibfnamefont{E.}~\bibnamefont{Mehler}},
  \bibinfo{author}{\bibfnamefont{F.}~\bibnamefont{Pitici}},
  \bibinfo{author}{\bibfnamefont{H.}~\bibnamefont{Weinstein}},
  \bibnamefont{and} \bibinfo{author}{\bibfnamefont{K.}~\bibnamefont{Schulten}},
  \bibinfo{journal}{Biophys.\ J.} \textbf{\bibinfo{volume}{74}},
  \bibinfo{pages}{1622} (\bibinfo{year}{1998}).

\bibitem[{\citenamefont{Vigil et~al.}(2001)\citenamefont{Vigil, Gallagher,
  Trewhella, and Garcia}}]{vigil:garcia:01}
\bibinfo{author}{\bibfnamefont{D.}~\bibnamefont{Vigil}},
  \bibinfo{author}{\bibfnamefont{S.~C.} \bibnamefont{Gallagher}},
  \bibinfo{author}{\bibfnamefont{J.}~\bibnamefont{Trewhella}},
  \bibnamefont{and} \bibinfo{author}{\bibfnamefont{A.~E.}
  \bibnamefont{Garcia}}, \bibinfo{journal}{Biophys.\ J.}
  \textbf{\bibinfo{volume}{80}}, \bibinfo{pages}{2082} (\bibinfo{year}{2001}).

\bibitem[{\citenamefont{Zuckerman}(2004)}]{zuckerman:04}
\bibinfo{author}{\bibfnamefont{D.~M.} \bibnamefont{Zuckerman}},
  \bibinfo{journal}{J.\ Phys.\ Chem. B} \textbf{\bibinfo{volume}{108}},
  \bibinfo{pages}{5127} (\bibinfo{year}{2004}).

\bibitem[{\citenamefont{Chen and Hummer}(2007)}]{gwei:hummer:07}
\bibinfo{author}{\bibfnamefont{Y.-G.} \bibnamefont{Chen}} \bibnamefont{and}
  \bibinfo{author}{\bibfnamefont{G.}~\bibnamefont{Hummer}},
  \bibinfo{journal}{J.\ Am.\ Chem.\ Soc.} \textbf{\bibinfo{volume}{129}},
  \bibinfo{pages}{2414} (\bibinfo{year}{2007}).

\bibitem[{\citenamefont{Bixon and Zwanzig}(1978)}]{bixon:zwanzig:78}
\bibinfo{author}{\bibfnamefont{M.}~\bibnamefont{Bixon}} \bibnamefont{and}
  \bibinfo{author}{\bibfnamefont{R.}~\bibnamefont{Zwanzig}},
  \bibinfo{journal}{J.\ Chem.\ Phys.} \textbf{\bibinfo{volume}{68}},
  \bibinfo{pages}{1896} (\bibinfo{year}{1978}).

\bibitem[{\citenamefont{Okazaki et~al.}(2006)\citenamefont{Okazaki, Koga,
  Takada, Onuchic, and Wolynes}}]{okazaki:wolynes:06}
\bibinfo{author}{\bibfnamefont{K.}~\bibnamefont{Okazaki}},
  \bibinfo{author}{\bibfnamefont{N.}~\bibnamefont{Koga}},
  \bibinfo{author}{\bibfnamefont{S.}~\bibnamefont{Takada}},
  \bibinfo{author}{\bibfnamefont{J.~N.} \bibnamefont{Onuchic}},
  \bibnamefont{and} \bibinfo{author}{\bibfnamefont{P.~G.}
  \bibnamefont{Wolynes}}, \bibinfo{journal}{Proc.\ Natl.\ Acad.\ Sci.\ USA}
  \textbf{\bibinfo{volume}{103}}, \bibinfo{pages}{11844}
  (\bibinfo{year}{2006}).

\bibitem[{\citenamefont{Best et~al.}(2005)\citenamefont{Best, Chen, and
  Hummer}}]{best:hummer:05}
\bibinfo{author}{\bibfnamefont{R.~B.} \bibnamefont{Best}},
  \bibinfo{author}{\bibfnamefont{Y.~G.} \bibnamefont{Chen}}, \bibnamefont{and}
  \bibinfo{author}{\bibfnamefont{G.}~\bibnamefont{Hummer}},
  \bibinfo{journal}{Structure} \textbf{\bibinfo{volume}{13}},
  \bibinfo{pages}{1755} (\bibinfo{year}{2005}).

\bibitem[{\citenamefont{Maragakis and Karplus}(2005)}]{maragakis:karplus:05}
\bibinfo{author}{\bibfnamefont{P.}~\bibnamefont{Maragakis}} \bibnamefont{and}
  \bibinfo{author}{\bibfnamefont{M.}~\bibnamefont{Karplus}},
  \bibinfo{journal}{J.\ Mol.\ Biol.} \textbf{\bibinfo{volume}{352}},
  \bibinfo{pages}{807} (\bibinfo{year}{2005}).

\bibitem[{\citenamefont{Miyazawa and Jernigan}(1996)}]{miyazawa:jernigan:96}
\bibinfo{author}{\bibfnamefont{S.}~\bibnamefont{Miyazawa}} \bibnamefont{and}
  \bibinfo{author}{\bibfnamefont{R.~L.} \bibnamefont{Jernigan}},
  \bibinfo{journal}{J.\ Mol.\ Biol.} \textbf{\bibinfo{volume}{256}},
  \bibinfo{pages}{623} (\bibinfo{year}{1996}).

\bibitem[{\citenamefont{Kim et~al.}(2002)\citenamefont{Kim, Jernigan, and
  Chirikjian}}]{kim:chirikjian:02}
\bibinfo{author}{\bibfnamefont{M.~K.} \bibnamefont{Kim}},
  \bibinfo{author}{\bibfnamefont{R.~L.} \bibnamefont{Jernigan}},
  \bibnamefont{and} \bibinfo{author}{\bibfnamefont{G.~S.}
  \bibnamefont{Chirikjian}}, \bibinfo{journal}{Biophys.\ J.}
  \textbf{\bibinfo{volume}{83}}, \bibinfo{pages}{1620} (\bibinfo{year}{2002}).

\bibitem[{\citenamefont{Russell and Barton}(1992)}]{russell:barton:92}
\bibinfo{author}{\bibfnamefont{R.~B.} \bibnamefont{Russell}} \bibnamefont{and}
  \bibinfo{author}{\bibfnamefont{G.~J.} \bibnamefont{Barton}},
  \bibinfo{journal}{Proteins Struct.\ Funct.\ Genet.}
  \textbf{\bibinfo{volume}{14}}, \bibinfo{pages}{309} (\bibinfo{year}{1992}).

\bibitem[{\citenamefont{Likic et~al.}(2003)\citenamefont{Likic, Strehler, and
  Gooley}}]{likic:gooley:03}
\bibinfo{author}{\bibfnamefont{V.~A.} \bibnamefont{Likic}},
  \bibinfo{author}{\bibfnamefont{E.~E.} \bibnamefont{Strehler}},
  \bibnamefont{and} \bibinfo{author}{\bibfnamefont{P.~R.}
  \bibnamefont{Gooley}}, \bibinfo{journal}{Protein Sci.}
  \textbf{\bibinfo{volume}{12}}, \bibinfo{pages}{2215} (\bibinfo{year}{2003}).

\bibitem[{\citenamefont{Strynadka and James}(1989)}]{strynadka:james:89}
\bibinfo{author}{\bibfnamefont{N.~C.~J.} \bibnamefont{Strynadka}}
  \bibnamefont{and} \bibinfo{author}{\bibfnamefont{M.~N.} \bibnamefont{James}},
  \bibinfo{journal}{Annu.\ Rev.\ Biochem.} \textbf{\bibinfo{volume}{58}},
  \bibinfo{pages}{951} (\bibinfo{year}{1989}).

\bibitem[{\citenamefont{Siivari et~al.}(1995)\citenamefont{Siivari, Zhang,
  Arthur G.~Palmer, and Vogel}}]{siivari:vogel:95}
\bibinfo{author}{\bibfnamefont{K.}~\bibnamefont{Siivari}},
  \bibinfo{author}{\bibfnamefont{M.}~\bibnamefont{Zhang}},
  \bibinfo{author}{\bibfnamefont{I.}~\bibnamefont{Arthur G.~Palmer}},
  \bibnamefont{and} \bibinfo{author}{\bibfnamefont{H.~J.} \bibnamefont{Vogel}},
  \bibinfo{journal}{FEBS Lett.} \textbf{\bibinfo{volume}{366}},
  \bibinfo{pages}{104} (\bibinfo{year}{1995}).

\bibitem[{\citenamefont{Shoemaker et~al.}(2000)\citenamefont{Shoemaker,
  Portman, and Wolynes}}]{shoemaker:wolynes:00}
\bibinfo{author}{\bibfnamefont{B.~A.} \bibnamefont{Shoemaker}},
  \bibinfo{author}{\bibfnamefont{J.~J.} \bibnamefont{Portman}},
  \bibnamefont{and} \bibinfo{author}{\bibfnamefont{P.~G.}
  \bibnamefont{Wolynes}}, \bibinfo{journal}{Proc.\ Natl.\ Acad.\ Sci.\ USA}
  \textbf{\bibinfo{volume}{97}}, \bibinfo{pages}{8868} (\bibinfo{year}{2000}).

\bibitem[{\citenamefont{Sinha et~al.}(2001)\citenamefont{Sinha, Kumar, and
  Nussinov}}]{sinha:nussinov:01}
\bibinfo{author}{\bibfnamefont{N.}~\bibnamefont{Sinha}},
  \bibinfo{author}{\bibfnamefont{S.}~\bibnamefont{Kumar}}, \bibnamefont{and}
  \bibinfo{author}{\bibfnamefont{R.}~\bibnamefont{Nussinov}},
  \bibinfo{journal}{Structure} \textbf{\bibinfo{volume}{9}},
  \bibinfo{pages}{1165} (\bibinfo{year}{2001}).

\bibitem[{\citenamefont{Lewit-Bentley and Rety}(2000)}]{bentley:rety:00}
\bibinfo{author}{\bibfnamefont{A.}~\bibnamefont{Lewit-Bentley}}
  \bibnamefont{and} \bibinfo{author}{\bibfnamefont{S.}~\bibnamefont{Rety}},
  \bibinfo{journal}{Curr.\ Opin.\ Struct.\ Biol.}
  \textbf{\bibinfo{volume}{10}}, \bibinfo{pages}{637} (\bibinfo{year}{2000}).

\bibitem[{\citenamefont{Lundstrom et~al.}(2005)\citenamefont{Lundstrom, Mulder,
  and Akke}}]{lundstrom:akke:05}
\bibinfo{author}{\bibfnamefont{P.}~\bibnamefont{Lundstrom}},
  \bibinfo{author}{\bibfnamefont{F.~A.~A.} \bibnamefont{Mulder}},
  \bibnamefont{and} \bibinfo{author}{\bibfnamefont{M.}~\bibnamefont{Akke}},
  \bibinfo{journal}{Proc.\ Natl.\ Acad.\ Sci.\ USA}
  \textbf{\bibinfo{volume}{102}}, \bibinfo{pages}{16984}
  (\bibinfo{year}{2005}).

\bibitem[{\citenamefont{Bryngelson et~al.}(1995)\citenamefont{Bryngelson,
  Onuchic, Socci, and Wolynes}}]{bryngelson:wolynes:95}
\bibinfo{author}{\bibfnamefont{J.~D.} \bibnamefont{Bryngelson}},
  \bibinfo{author}{\bibfnamefont{J.~N.} \bibnamefont{Onuchic}},
  \bibinfo{author}{\bibfnamefont{N.~D.} \bibnamefont{Socci}}, \bibnamefont{and}
  \bibinfo{author}{\bibfnamefont{P.~G.} \bibnamefont{Wolynes}},
  \bibinfo{journal}{Proteins Struct.\ Funct.\ Genet.}
  \textbf{\bibinfo{volume}{21}}, \bibinfo{pages}{167} (\bibinfo{year}{1995}).

\bibitem[{\citenamefont{Verkhivker et~al.}(2002)\citenamefont{Verkhivker,
  Bouzida, Gehlhaar, Rejto, Freer, and Rose}}]{verkhivker:rose:02}
\bibinfo{author}{\bibfnamefont{G.~M.} \bibnamefont{Verkhivker}},
  \bibinfo{author}{\bibfnamefont{D.}~\bibnamefont{Bouzida}},
  \bibinfo{author}{\bibfnamefont{D.~K.} \bibnamefont{Gehlhaar}},
  \bibinfo{author}{\bibfnamefont{P.~A.} \bibnamefont{Rejto}},
  \bibinfo{author}{\bibfnamefont{S.~T.} \bibnamefont{Freer}}, \bibnamefont{and}
  \bibinfo{author}{\bibfnamefont{P.~W.} \bibnamefont{Rose}},
  \bibinfo{journal}{Curr.\ Opin.\ Struct.\ Biol.}
  \textbf{\bibinfo{volume}{12}}, \bibinfo{pages}{197} (\bibinfo{year}{2002}).

\bibitem[{\citenamefont{Wang et~al.}(2006)\citenamefont{Wang, Zhang, Lu, and
  Wang}}]{wang:wang:06}
\bibinfo{author}{\bibfnamefont{J.}~\bibnamefont{Wang}},
  \bibinfo{author}{\bibfnamefont{K.}~\bibnamefont{Zhang}},
  \bibinfo{author}{\bibfnamefont{H.~Y.} \bibnamefont{Lu}}, \bibnamefont{and}
  \bibinfo{author}{\bibfnamefont{E.~K.} \bibnamefont{Wang}},
  \bibinfo{journal}{Phys.\ Rev.\ Lett.} \textbf{\bibinfo{volume}{96}},
  \bibinfo{pages}{168101} (\bibinfo{year}{2006}).

\bibitem[{\citenamefont{Papoian and Wolynes}(2003)}]{papoian:wolynes:03}
\bibinfo{author}{\bibfnamefont{G.~A.} \bibnamefont{Papoian}} \bibnamefont{and}
  \bibinfo{author}{\bibfnamefont{P.}~\bibnamefont{Wolynes}},
  \bibinfo{journal}{Biopolymers} \textbf{\bibinfo{volume}{63}},
  \bibinfo{pages}{333} (\bibinfo{year}{2003}).

\bibitem[{\citenamefont{Verkhivker et~al.}(2003)\citenamefont{Verkhivker,
  Bouzida, Gellhaar, Tejto, Freer, and Rose}}]{verkhivker:rose:03}
\bibinfo{author}{\bibfnamefont{G.~M.} \bibnamefont{Verkhivker}},
  \bibinfo{author}{\bibfnamefont{D.}~\bibnamefont{Bouzida}},
  \bibinfo{author}{\bibfnamefont{D.~K.} \bibnamefont{Gellhaar}},
  \bibinfo{author}{\bibfnamefont{P.~A.} \bibnamefont{Tejto}},
  \bibinfo{author}{\bibfnamefont{S.~T.} \bibnamefont{Freer}}, \bibnamefont{and}
  \bibinfo{author}{\bibfnamefont{P.~W.} \bibnamefont{Rose}},
  \bibinfo{journal}{Proc.\ Natl.\ Acad.\ Sci.\ USA}
  \textbf{\bibinfo{volume}{100}}, \bibinfo{pages}{5148} (\bibinfo{year}{2003}).

\bibitem[{\citenamefont{Miyashita et~al.}(2003)\citenamefont{Miyashita,
  Onuchic, and Wolynes}}]{miyashita:wolynes:03}
\bibinfo{author}{\bibfnamefont{O.}~\bibnamefont{Miyashita}},
  \bibinfo{author}{\bibfnamefont{J.~N.} \bibnamefont{Onuchic}},
  \bibnamefont{and} \bibinfo{author}{\bibfnamefont{P.~G.}
  \bibnamefont{Wolynes}}, \bibinfo{journal}{Proc.\ Natl.\ Acad.\ Sci.\ USA}
  \textbf{\bibinfo{volume}{100}}, \bibinfo{pages}{12570}
  (\bibinfo{year}{2003}).

\bibitem[{\citenamefont{Levy et~al.}(2004)\citenamefont{Levy, Wolynes, and
  Onuchic}}]{levy:onuchic:04}
\bibinfo{author}{\bibfnamefont{Y.}~\bibnamefont{Levy}},
  \bibinfo{author}{\bibfnamefont{P.~G.} \bibnamefont{Wolynes}},
  \bibnamefont{and} \bibinfo{author}{\bibfnamefont{J.~N.}
  \bibnamefont{Onuchic}}, \bibinfo{journal}{Proc.\ Natl.\ Acad.\ Sci.\ USA}
  \textbf{\bibinfo{volume}{101}}, \bibinfo{pages}{511} (\bibinfo{year}{2004}).

\bibitem[{\citenamefont{Yang et~al.}(2004)\citenamefont{Yang, Cho, Levy,
  Cheung, Levine, Wolynes, and Onuchic}}]{yang:onuchic:04}
\bibinfo{author}{\bibfnamefont{S.}~\bibnamefont{Yang}},
  \bibinfo{author}{\bibfnamefont{S.~S.} \bibnamefont{Cho}},
  \bibinfo{author}{\bibfnamefont{Y.}~\bibnamefont{Levy}},
  \bibinfo{author}{\bibfnamefont{M.~S.} \bibnamefont{Cheung}},
  \bibinfo{author}{\bibfnamefont{H.}~\bibnamefont{Levine}},
  \bibinfo{author}{\bibfnamefont{P.~G.} \bibnamefont{Wolynes}},
  \bibnamefont{and} \bibinfo{author}{\bibfnamefont{J.~N.}
  \bibnamefont{Onuchic}}, \bibinfo{journal}{Proc.\ Natl.\ Acad.\ Sci.\ USA}
  \textbf{\bibinfo{volume}{101}}, \bibinfo{pages}{13786}
  (\bibinfo{year}{2004}).

\bibitem[{\citenamefont{Levy et~al.}(2005)\citenamefont{Levy, Cho, Onuchic, and
  Wolynes}}]{levy:wolynes:05}
\bibinfo{author}{\bibfnamefont{Y.}~\bibnamefont{Levy}},
  \bibinfo{author}{\bibfnamefont{S.~S.} \bibnamefont{Cho}},
  \bibinfo{author}{\bibfnamefont{J.~N.} \bibnamefont{Onuchic}},
  \bibnamefont{and} \bibinfo{author}{\bibfnamefont{P.~G.}
  \bibnamefont{Wolynes}}, \bibinfo{journal}{J.\ Mol.\ Biol.}
  \textbf{\bibinfo{volume}{346}}, \bibinfo{pages}{1121} (\bibinfo{year}{2005}).

\bibitem[{\citenamefont{Huang and Montelione}(2005)}]{huang:montelione:05}
\bibinfo{author}{\bibfnamefont{Y.~J.} \bibnamefont{Huang}} \bibnamefont{and}
  \bibinfo{author}{\bibfnamefont{G.~T.} \bibnamefont{Montelione}},
  \bibinfo{journal}{Nature} \textbf{\bibinfo{volume}{438}}, \bibinfo{pages}{36}
  (\bibinfo{year}{2005}).

\bibitem[{\citenamefont{Tozzini}(2005)}]{tozzini:05}
\bibinfo{author}{\bibfnamefont{V.}~\bibnamefont{Tozzini}},
  \bibinfo{journal}{Curr.\ Opin.\ Struct.\ Biol.}
  \textbf{\bibinfo{volume}{15}}, \bibinfo{pages}{144} (\bibinfo{year}{2005}).

\bibitem[{\citenamefont{Tirion}(1996)}]{tirion:96}
\bibinfo{author}{\bibfnamefont{M.~M.} \bibnamefont{Tirion}},
  \bibinfo{journal}{Phys. Rev. Lett.} \textbf{\bibinfo{volume}{77}},
  \bibinfo{pages}{1905} (\bibinfo{year}{1996}).

\bibitem[{\citenamefont{Bahar et~al.}(1998)\citenamefont{Bahar, Atilgan,
  Demirel, and Erman}}]{bahar:erman:98}
\bibinfo{author}{\bibfnamefont{I.}~\bibnamefont{Bahar}},
  \bibinfo{author}{\bibfnamefont{A.~R.} \bibnamefont{Atilgan}},
  \bibinfo{author}{\bibfnamefont{M.~C.} \bibnamefont{Demirel}},
  \bibnamefont{and} \bibinfo{author}{\bibfnamefont{B.}~\bibnamefont{Erman}},
  \bibinfo{journal}{Phys.\ Rev.\ Lett.} \textbf{\bibinfo{volume}{80}},
  \bibinfo{pages}{2733} (\bibinfo{year}{1998}).

\bibitem[{\citenamefont{Tama and Brooks}(2002)}]{tama:brooks:02}
\bibinfo{author}{\bibfnamefont{F.}~\bibnamefont{Tama}} \bibnamefont{and}
  \bibinfo{author}{\bibfnamefont{C.~L.} \bibnamefont{Brooks}},
  \bibinfo{journal}{J.\ Mol.\ Biol.} \textbf{\bibinfo{volume}{318}},
  \bibinfo{pages}{733} (\bibinfo{year}{2002}).

\bibitem[{\citenamefont{Miyashita et~al.}(2004)\citenamefont{Miyashita,
  Wolynes, and Onuchic}}]{miyashita:wolynes:04}
\bibinfo{author}{\bibfnamefont{O.}~\bibnamefont{Miyashita}},
  \bibinfo{author}{\bibfnamefont{P.~G.} \bibnamefont{Wolynes}},
  \bibnamefont{and} \bibinfo{author}{\bibfnamefont{J.~N.}
  \bibnamefont{Onuchic}}, \bibinfo{journal}{J.\ Phys.\ Chem. B}
  \textbf{\bibinfo{volume}{109}}, \bibinfo{pages}{1959} (\bibinfo{year}{2004}).

\bibitem[{\citenamefont{Zong et~al.}(2007)\citenamefont{Zong, Wilson, Shen,
  Wittung-Stafshede, Mayo, and Wolynes}}]{zong:wolynes:07}
\bibinfo{author}{\bibfnamefont{C.}~\bibnamefont{Zong}},
  \bibinfo{author}{\bibfnamefont{C.~J.} \bibnamefont{Wilson}},
  \bibinfo{author}{\bibfnamefont{T.}~\bibnamefont{Shen}},
  \bibinfo{author}{\bibfnamefont{P.}~\bibnamefont{Wittung-Stafshede}},
  \bibinfo{author}{\bibfnamefont{S.~L.} \bibnamefont{Mayo}}, \bibnamefont{and}
  \bibinfo{author}{\bibfnamefont{P.~G.} \bibnamefont{Wolynes}},
  \bibinfo{journal}{Proc.\ Natl.\ Acad.\ Sci.\ USA}
  \textbf{\bibinfo{volume}{104}}, \bibinfo{pages}{3159} (\bibinfo{year}{2007}).

\bibitem[{\citenamefont{Humphrey et~al.}(1996)\citenamefont{Humphrey, Dalke,
  and Schulten}}]{humphrey:schulten:96}
\bibinfo{author}{\bibfnamefont{W.}~\bibnamefont{Humphrey}},
  \bibinfo{author}{\bibfnamefont{A.}~\bibnamefont{Dalke}}, \bibnamefont{and}
  \bibinfo{author}{\bibfnamefont{K.}~\bibnamefont{Schulten}},
  \bibinfo{journal}{J.\ Mol.\ Graphics} \textbf{\bibinfo{volume}{14}},
  \bibinfo{pages}{33} (\bibinfo{year}{1996}).

\end{thebibliography}

\clearpage

\section*{Figure Legends}
\subsubsection*{Figure~\ref{fig:2nd_3D}.}

The N-terminal domain of calmodulin (nCaM). (a) The
Ca$^{2+}$-free (apo, closed) structure, PDB code 1cfd. (b) The
Ca$^{2+}$-bound (holo, open) structure, PDB code 1cll. (c) The
secondary structure of nCaM is shown with one letter amino acid
sequence code for residues 4-75. The secondary structure of nCaM is as
follows: helix A (5--19), Ca$^{2+}$-binding loop I (20--31), helix B
(29--37), B/C helix-linker (38--44), helix C (45--55), Ca$^{2+}$-binding
loop II (56--67), helix D (65--75). Note that, the last three residues 
of the binding loops I and II are also part of the exiting helices B 
and D. There are short $\beta$-sheet structures in binding loop I 
(residues 26--28) and loop II (residues 62--64). This, and other 
three-dimensional illustrations were made using Visual Molecular 
Dynamics (VMD).\cite{humphrey:schulten:96}

\subsubsection*{Figure~\ref{fig:Gii}.}

Fluctuations $B_i = \langle\delta\mathbf{r}_i^2\rangle_0 =
G_{ii}a^2$ vs sequence index of nCaM for selected values of the interpolation
parameter $\alpha_0$ in the conformational transition route between
open and closed. Here $a = 3.8$\AA
is the distance between successive
monomers.  Different $\alpha_0$ are denoted by, red ($\alpha_0=0$)
open; green ($\alpha_0=0.2$); blue ($\alpha_0=0.4$); pink
($\alpha_0=0.6$); orange ($\alpha_0=0.8$) and black ($\alpha_0=1$)
closed. The secondary structure is indicated below the plot.
\subsubsection*{Figure~\ref{fig:B_3D}.}
Change in fluctuations $B_i$ in nCaM domain during the
closed to open conformational transition. The 3D structure in (a)
corresponds to the interpolation parameter, $\alpha_0=1$ (closed
state); (b) corresponds to $\alpha_0=0.4$ (intermediate state) and (c)
corresponds to $\alpha_0=0$ (open state). Red corresponds to low
fluctuations and blue corresponds to high. Here, $a$ is the distance
between successive monomers.

\subsubsection*{Figure~\ref{fig:drhon}.}

Difference between the normalized native density
$\Delta\overline{\rho_i}$ (a measure of structural similarity) of each
residue for different $\alpha_0$. The change in color from red to blue
is showing the closed $\rightarrow$ open conformational transition of
nCaM. This is normalized to be $-1$ at the open state minimum
($\alpha_0=0$; blue) and 1 at the closed state minimum ($\alpha_0=1$;
red). Below the secondary structure of nCaM is shown. Here, 
$\alpha^N$ in Eq.~\ref{eq:density 1} and Eq.~\ref{eq:density 2} is 0.5.
\subsubsection*{Figure~\ref{fig:drhon_3D}.}
Closed to open conformational transition in nCaM with
different interpolation parameter $\alpha_0$. The 3D structure in (a)
corresponds to the interpolation parameter, $\alpha_0=0.8$; (b)
corresponds to $\alpha_0=0.6$; (c) corresponds to $\alpha_0=0.4$ and
(d) corresponds to $\alpha_0=0.2$. The change in color from red to
blue corresponds to different values of normalized native density
$\Delta\overline{\rho_i}$ (a measure of structural similarity) of each
residue for different $\alpha_0$. Red corresponds to
$\Delta\overline{\rho_i}=1$ (closed conformation) and blue (open
conformation) corresponds to $\Delta\overline{\rho_i}= -1$.

\subsubsection*{Figure~\ref{fig:loops_3D}.}

Comparison of structural change in binding loops I 
(in bottom) and II (in top) in terms of the order parameter 
$\Delta\overline{\rho_i}$. The 3D structures in (a)-(i) 
corresponds to the interpolation parameter, $\alpha_0=0.9$
-0.1 during the closed to open transition. The change in 
color from red to blue corresponds to different values of 
$\Delta\overline{\rho_i}$ (a measure of structural similarity) 
of each residue. Red corresponds to $\Delta\overline{\rho_i}=1$
(closed conformation) and blue (open conformation) corresponds
to $\Delta\overline{\rho_i}= -1$.

\subsubsection*{Figure~\ref{fig:residues}.}

Dynamical behavior of residues during conformational
transition of nCaM. The normalized native density difference
$\Delta\overline{\rho_i}$ vs $\alpha_0$ are shown for four different
group of residues. Structural transition of (a) residues in position 9 
(Thr28 and Asn64) and position 12 (Glu31 and Glu67) of the two binding 
loops; (b) four hydrophobic Methionine residues in positions 36, 51, 71 and 72.

\begin{figure}
   \begin{center}
      \includegraphics[width=5.0in]{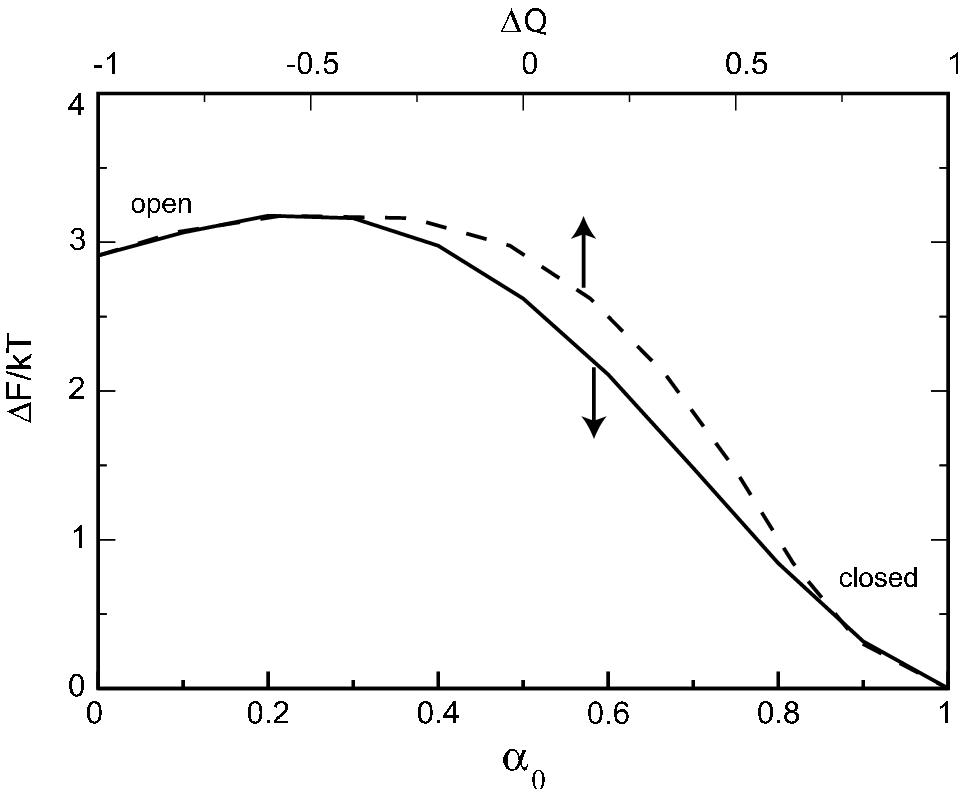}
      \caption{}
      \label{fig:f_Q}
   \end{center}
\end{figure}

\subsubsection*{Figure~\ref{fig:f_Q}.}

Free energy along the transition route. In the lower
curve the abscissa is the interpolation parameter
$\alpha_0$. In the upper curve the abscissa is the global structural
order parameter $\Delta Q$. The entropy across the transition is 
relatively constant, so that the free energy barrier is largely energetic.

\clearpage

\clearpage

\clearpage

\clearpage

\clearpage

\clearpage

\clearpage

\end{document}